\documentclass[aps,prd,epsf,amsmath,amssymb, 11pt]{revtex4}
\usepackage{color,graphicx}
\usepackage{dcolumn}% Align table columns on decimal point
\usepackage{bm}% bold math
\usepackage{mathrsfs}
\usepackage[T1]{fontenc}

\newcommand{\be}{\begin{equation}}
\newcommand{\ee}{\end{equation}}
\newcommand{\ba}{\begin{eqnarray}}
\newcommand{\ea}{\end{eqnarray}}
\newcommand{\ban}{\begin{eqnarray*}}
\newcommand{\ean}{\end{eqnarray*}}

\def\dbar{{\mathchar'26\mkern-12mu d}}

\begin{document}

\title{A new proposal regarding the heat generated by gravity in locally accelerating frames}

\author{Swastik Bhattacharya}\email{swastik@imsc.res.in}
\affiliation{ The Institute of Mathematical Sciences, 
Chennai, India}

\begin{abstract} 
 For Rindler observers accelerating close  to the horizon in local patches around a spacetime point,
 the matter-energy passing through the horizon increases the entropy and heat energy. Jacobson has showed 
 that the Einstein  equation can be derived from the consideration of this thermodynamic process. This, 
 however, works only if the 
 acceleration $a$ is much larger than the scale set by  the curvature of the spacetime. It is explored here 
 whether an extension is possible to the case with no lower bound on $a$. We show that this is 
 possible if one assumes that in a locally accelerating frame, the matter-energy passing through null 
 hypersurfaces could result in an increase in the heat energy and the entropy. Such a generalisation extends 
 the thermodynamic derivation of gravity to include any non-freely falling observer. A new method of 
 determining the temperature detected by such locally 
 accelerating observers is also presented. By considering only the quantisation of sufficiently localised 
wave modes of a field, it is shown that the observer finds himself in a thermal environment.

%  Assuming that the entropy is proportional to the area of the locally defined Rindler horizon around a spacetime point, Jacobson 
% has pointed out that the Einstein equation follows from Thermodynamic considerations. Here it is shown that this result can be 
% generalised for observers accelerating locally around a spacetime point. One can consider an observer who is accelerating in an 
% appropriately defined local patch around some spacetime point. It can be shown that under some reasonable physical assumptions, such
% an observer would detect thermal radiation. Also, following a line of reasoning similar to that of Jacobson's, the Einstein equation 
% can be derived from Thermodynamic considerations in this case.  
% . 

\end{abstract}
%\pacs{04.20.Dw, 04.70.-s, 04.70.Bw}
\maketitle

\section{Introduction}
There are strong reasons to believe that there is a connection between Gravity and Thermodynamics, even 
though, so far we do not know what the microscopic degrees of freedom are in this case. 
Historically, the laws of Black hole mechanics \cite{BHTherm}, which are of a form strikingly 
similar to that of the laws of thermodynamics, came first. Motivated by this and other considerations, Bekenstein 
proposed that the Black holes have a temperature and possess an entropy proportional to the surface area of 
the Black hole event horizon\cite{Bekenstein}. The similarity became much more physically compelling, once Hawking 
radiation was discovered from semiclassical considerations\cite{Hawking}. Around the same time, it was 
discovered that even in the Minkowski spacetime, an accelerating observer in a Rindler frame would encounter 
thermal radiation whereas an inertial observer sees only vacuum \cite{Unruh}.

% 
%  But locally any 
% spacetime can be viewed as flat. Now let us construct an 'inertial' frame in a local patch around a 
% particular spacetime point. This means that in the coordinate system(C.S.) locally the metric is of 
% Minkowskian form. Let us further assume that any 'inertial' observer in such a patch finds a field to be in a
% vacuum state. Then it follows that any 'Rindler observer in this local patch' would detect thermal 
% fluctuations. For this to work, one has to consider a local patch small enough such that the curvature 
% effects can be neglected. Typically, within a small enough local patch, the curvature effects could be 
% neglected up to first order, e.g. in a Riemann normal C.S. set up around a spacetime point. 
%In other words, the curvature effects cannot be larger than second order in magnitude.)* \\
% *(It also turns out that black holes and Rindler like horizons both behave similarly when matter-energy 
% passes through them. Some version of the laws of thermodynamics hold for both of them(?)[see Jacobson, 
% Parentini for discussion on this] )*\\

The black hole spacetime and the Rindler spacetime have one thing in common, causal horizons exist for both 
of them. The vacuum fluctuations in the frame of the inertial observer are seen as thermal fluctuations 
by the Rindler observer. This correspondence holds for the flat spacetime. However, it is 
possible to consider accelerating observers at any spacetime point and construct a locally Rindler 
patch around it. Such observers are expected to find themselves in a thermal environment. The temperature 
detected by them is proportional to 
the acceleration as in Rindler spacetime. One notices here an interesting similarity between observers 
who remain outside black holes and observers who are locally accelerating. For black holes, however, the 
analogy of the laws of black hole mechanics with thermodynamics was noticed before the discovery of the 
thermal nature of the radiation from the black hole. But it is also possible to proceed in the reverse 
direction and explore whether the thermal nature of the radiation encountered locally by the accelerated observer in a
spacetime is accompanied by some interesting consequences of thermodynamics. A pioneering step in this direction 
was taken by Jacobson\cite{Jacobson}.  For this, he had to use the thermodynamic properties of a 
Rindler-like horizon in a local construction. So he considered an observer accelerating at a constant rate 
in an arbitrary spacetime in such a way that one can construct a locally Rindler patch containing the observer, 
with the Rindler horizon forming a part of its boundary. The amount of matter-energy that passed through this horizon was taken to be 
the increase in the heat energy, $\dbar Q$, while the increase in the horizon area was assumed to be proportional 
the increase in the entropy $ds$ of the system. Assuming this, he was able to show that the relation, $\dbar Q = Tds$, 
then gives back the 
Einstein equation. Based on this result, it was suggested by him that gravity itself could be viewed as an emergent phenomenon 
[Also see discussions in \cite{Paddy}]. This derivation has recently been extended to include some other theories of 
gravity as well \cite{Eling}, \cite{Chirco}, \cite{Silva}, \cite{Hadad}, \cite{Sudipta}, \cite{Arif}.

It is clear that such a derivation of the Einstein equation has two inherent limitations. The
first one is that for the inertial-Rindler observer correspondence between vacuum-thermal fluctuations to 
work, the spacetime has to be flat and this in general is only possible to achieve locally. That, however, does not pose any real 
problem. In fact, it could be thought of as an attractive feature of this derivation, where the Einstein equation 
emerges as a consequence of the thermodynamics at a local level.

The second limitation is essentially a restriction in the class of locally accelerating observers, for whom 
this derivation of the Einstein equation from thermodynamics works. It arises mainly from the consideration 
of 
what the accelerating observer is going to see and interpret as an increase in heat energy. The idea here is 
that the information about the degrees of freedom of a macroscopic body is lost to the observer as it reaches 
the horizon. That energy then appears as heat energy to the observer, who remains  outside the horizon. This 
intuition receives strong support from other considerations like the first law of horizon thermodynamics \cite{JacobParent}
 and the results obtained within the membrane paradigm for black holes \cite{membrane}. 
This implies that to make use of the thermodynamic correspondence, the local Rindler patch has to be 
constructed in such a way that the Rindler horizon forms a part of the boundary of that patch. In addition to 
this, there is a further constraint, which is more technical in nature. For Jacobson's derivation to work, 
one has to consider matter-energy passing through close to the bifurcation surface of the Rindler horizon of 
the local Rindler patch. The bifurcation surface of a Rindler wedge is at a distance of  
$\frac{1}{a}$ from the accelerating observer. Now, from the consideration of Riemann normal C.S., the size $L$ 
of the local patch would be such that $L\ll \frac{1}{\sqrt{Riemann}}$. The distance of the Rindler observer 
from the bifurcation point should not exceed that length scale. It follows then that the acceleration should 
be such that, $\frac{1}{a}\ll \frac{1}{\sqrt{Riemann}}$. This condition restricts the class of locally 
accelerating observers, for whom the Einstein equation is a consequence of thermodynamics. Thus among all 
the observers, who are locally accelerating, the above derivation of Einstein equation from thermodynamics 
would only hold for the subclass of observers, who accelerate at a rate much higher than the scale set by 
the curvature of the spacetime.  It is possible to take a limit where the acceleration goes to infinity and 
so does the boost energy, which passes through the horizon, making the analysis strictly local. In fact, this 
was the way, that this analysis was carried out originally by Jacobson \cite{Jacobson}. From the thermodynamic point 
of view, this means that the temperature is going to infinity and so does the amount of heat generated as seen by the 
observer. There ratio remains finite however and gives the change in the entropy of the system. It is 
important to note here that one can take this limit sensibly because the amount of entropy increase is 
independent of the scale of the acceleration. This is a feature that is expected to 
persist in our generalisation of Jacobson's derivation.

The accelerating observer sees thermal radiation at a temperature proportional to $a$ in a Rindler patch bounded on one side by a 
Rindler horizon. It is less clear however, whether an observer accelerating locally, for whom the Rindler horizon does not 
exist in a local patch, would still find herself in a thermal environment. Here we demonstrate by a semiclassical 
calculation that such an observer would detect a temperature proportional to the 
acceleration. We focus on the Fourier modes of a massless scalar field, which are localised within this 
patch and build up Fock states using only those modes. Then it can be shown that provided the 'inertial 
observer' sees a vacuum state, the accelerating observer finds the environment to be thermal with a 
temperature proportional to the acceleration.

Normally, if a system has a temperature, then one can reasonably expect that a thermodynamic description 
exists for the macroscopic processes concerning the system. This was found to be true for black holes. 
Jacobson's derivation of Einstein equation from thermodynamics is also a confirmation of that expectation. From 
this point of view, the restriction of this derivation to a subclass of locally accelerating observers 
appears unsatisfactory. The locally accelerating observers who have been excluded, also detect a temperature. 
 So a thermodynamic description should exist for all the observers, who are accelerating locally.

One can also note the work done by Verlinde\cite{Verlinde} in this context, who claimed that gravity can be thought of 
as an entropic force. In particular, assuming that the relevant degrees of freedom for gravity resides on two dimensional 
spacelike surfaces, through which the matter passes and  Equipartition theorem holds for such degrees of freedom, he was able to 
derive the Einstein equation, or more precisely, the time-time component of it \cite{Miao}. Our approach here is different from the 
entropic gravity perspective and hence we would not refer to it any further. However, it is important to note 
that his work suggested that a thermodynamic interpretation may also be possible for observers not belonging to the class, 
for which Jacobson's method works.

The possibility of extending the thermodynamic derivation of the Einstein equation to all observers who are 
accelerating locally would be explored here. For this, one needs a prescription to determine both the 
increase in entropy and heat energy. To be able to do this, a geometric structure has to be identified, to 
which, one could assign thermodynamic properties. In the case considered by Jacobson, the Rindler horizon served this 
purpose. However, locally one can always find a class of accelerating observers, who see a given null hypersurface 
as the Rindler horizon.  Moreover, as Padmanabhan and Paranjape has showed \cite{Paddy1}, one can derive the 
dynamical equations of gravity by 
assigning an entropy functional to such a null hypersurface. In fact, using the Einstein equation, it is 
possible to write down the dynamics of such null hypersurfaces in the form of Navier-Stokes equation
\cite{Damour1},\cite{Damour2},\cite{Paddy2},\cite{Paddy3}.
All this suggests that such null hypersurfaces are the natural choice for the generalisation sought here. 
Accordingly, we choose the null hypersurfaces within the local patch to determine the increase in entropy and heat. 
The change in entropy is assumed to be proportional to the change in the area of the cross section of the 
null boundary of the local patch. This may seem somewhat similar in spirit to the estimation of the degrees of 
freedom in a causal diamond according to the Holographic principle\cite{holography}.

How to determine the increase in the amount of heat energy is less apparent, however. A proposal is made here, 
which leads to a prescription, that could be used to compute $\dbar Q$. 
The main idea here is that a fraction of the matter-energy passing through a null hypersurface gets converted
into heat in a locally accelerating frame. When the null surface is not a horizon, the fraction of 
matter-energy that gets converted would be typically very small. Two criteria have been used to arrive at such a
prescription to determine $\dbar Q$. \\
1) When evaluated using this prescription, $\dbar Q = Tds$ should give the Einstein equation. \\ 
2) When used to compute $\dbar Q$ for the matter-energy passing through the horizon for a locally Rindler-like 
observer, the prescription should be able to produce the standard result. \\
The main point in our paper then is the proposal that just like matter-energy passing through the horizon becomes 
heat energy and increases the entropy, in an accelerating frame, matter-energy passing through the 
null hypersurfaces within a local patch, could also result in the increase of heat and entropy
by a tiny amount.

The paper is organised as follows. In the next section, the Einstein equation is derived from thermodynamic 
considerations. In the third section, we show that how a person at rest in a locally accelerating frame 
sees a thermal environment. In the final section, the implications of this 
result are discussed.

\section{Derivation of the Einstein equation from Equilibrium Thermodynamics}

We now demonstrate how the Einstein equation can be derived from thermodynamic considerations. For this, we 
have to choose a local neighbourhood around some spacetime point $P$ in an arbitrary spacetime. After defining 
the local patch properly and then introducing an inertial and an accelerating coordinate system covering it, 
we can proceed with the thermodynamic analysis.

\subsection{Construction of a local patch and a causal diamond around a spacetime point}
\begin{figure}
\begin{center}
\includegraphics[width=8cm]{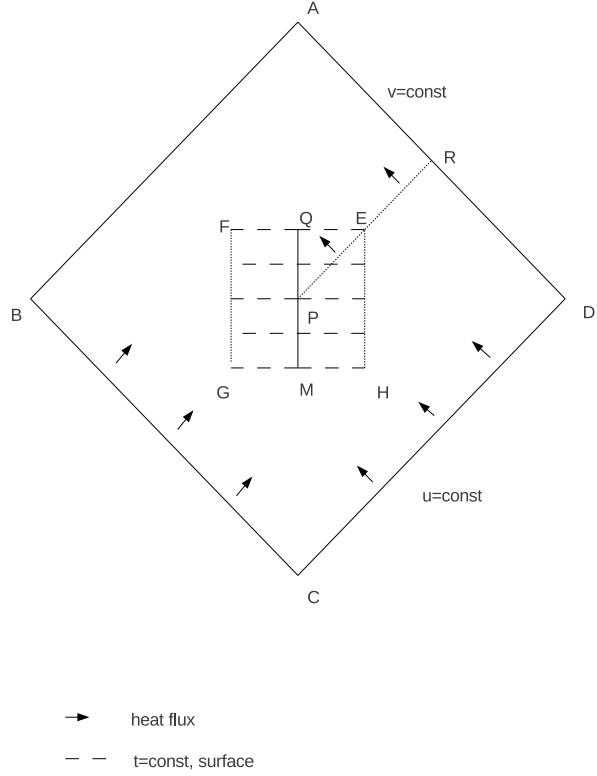}
\caption{\label{fg1} 
Locally constructed Causal Diamond $\mathscr{D}$ in the $t-n$ space}
\end{center}
\end{figure}

 We can now consider a causal diamond $\mathscr{D}$ (see FIG. \ref{fg1}) around the point $P$, sufficiently 
small, so that one can neglect the effects of the curvature of the spacetime within it. One can construct a C.S. with a 
Minkowski metric in this patch. Anybody at rest or in uniform motion in this C.S. can be thought of as an 
Inertial Observer.  Now we consider an observer moving with a constant acceleration $a$ in this local 
patch. One can also construct a C.S. within this causal diamond, in which the 
accelerating observer is at rest. We now choose the boundaries of $\mathscr{D}$ 
to be the null boundaries of the local patch, where the observer resides. This can be done in the 
following way. First, let us consider the local observer to be such that he or she could typically measure 
lengths or time differences much smaller than the size of the local patch. Suppose that the size of 
$\mathscr{D}$ is of order 
$L$. Then the length scale associated with the local observers would be much smaller than $L$. This can be achieved 
by rescaling the coordinates by a factor $\Lambda$ for the local observers such that $\Lambda\gg1$. This means that the 
size of $\mathscr{D}$ is now of the order of $\Lambda L$ in the rescaled units of the local observers. If one takes 
$\Lambda$ to be sufficiently large, 
$\Lambda L$ would also be a very large quantity. In that case, one may think of the size of the local patch to be as 
large as possible or practically infinity as far as the local observers are concerned. One can take only such 
null hypersurfaces as the null boundary of this local patch, which are themselves at a distance of order 
$\Lambda L$ away from any point in it. The choice is also made in such a way that the boundaries are 
null geodesics.

Such a construction is shown in FIG. \ref{fg1}. It represents only the $t-n$ part and the other two space dimensions have been
suppressed. $EFGH$ denotes the local patch around the point $P$. Let, $EF= GH = fL$ and $AC= BD= L$, where $f<1$ but 
$fL$ and $L$ are both of the same order. This means that the local Minkowskian patch can be extended to cover 
the whole of $\mathscr{D}$. $MPQ$ denotes 
the world line of an observer locally accelerating with acceleration $a$. One can draw null geodesics from any point in this 
patch, which intersects the causal diamond $ABCD$ surrounding it. $PER$ is such a null geodesic. The affine distance along 
such null geodesics from any point in $EFGH$ to $ABCD$ is of the order of $L$. Measured in the scale of the 
local observer, this distance would be very large. In other words, for a local observer within $EFGH$, the boundary of the
causal diamond $\mathscr{D}$ is very far away. So $\mathscr{D}$ can be taken as the null boundary of such an observer. 
Admittedly, such a choice of the null boundary is arbitrary to an extent. There is however an upper limit to the arbitrariness 
as the size of $\mathscr{D}$ has to be much less than the scale fixed by the curvature of the spacetime. So two such choices 
of the null boundary cannot be removed from each other by a distance greater than the order $L$.

% For convenience, we shall first state the main assumptions made by us and also 
% discuss the motivation behind them. \\
 
% (However, for our purposes, 
% we need to consider the observer who is at rest in an accelerating frame $A$. We would be
% making a few very important assumptions about the C.S. in the $A$ frame. They are discussed 
% below.)
%  Now we consider an observer moving with a constant acceleration $a$ within the local patch, 
% that we are considering. 
The velocity vector field of the accelerating observers, is denoted by $t^\mu$, where $t^\mu t_\mu=-1$. Let, 
the unit vector along the direction of acceleration be denoted by $n_\mu$.  Such an
observer is at rest in the accelerating frame $\mathcal{A}$. If we denote the coordinates in this frame by $(x^0, x^1, x^2, x^3)$, 
the energy flux 3-vector is $T^{0i}$, where $T_{\mu \nu}$ is the energy-momentum tensor. 
Now, in the frame $\mathcal{A}$, $t^a =0$, where $a$ denotes the spacelike indices. 
% {\bf(One can make a further 
% transformation in the 3 space dimensions such that $T^{0i}$ is of the form, $(T^{0n}, 0, 0)$. The direction 
% along which the energy flux vector has a non-zero component 
% is denoted by $n$. One can introduce a unit vector along this direction, $n_\mu$. This vector would be 
% spacelike, hence $n^\mu n_\mu= 1$. This construction allows us to view matter passing 
% through this local patch in the following way. One can imagine a little box in the three volume 
% enclosed by the local patch at the point $P$. Within the local patch, this box would evolve 
% in time. The energy flux passes through this box along the vector $n_\mu$.)} 

Since one can neglect the curvature effects inside the causal diamond $\mathscr{D}$, a timelike Killing 
vector can be defined there. We further demand that the C.S. in $A$ frame is such that there exists a timelike 
Killing vector, $\xi^\mu$, parallel to $t^{\mu}$. Also $\xi^\mu$ can be written as $(1, 0,0,0)$ and it 
satisfies the relation
\begin{equation}
 \xi^\mu_{;\nu} n^\nu = a \xi^\mu . \label{Killingchange}
\end{equation}
A priori, it is not obvious whether such a condition could be made to satisfy around any spacetime point $P$.
However, we are going to show later that this is indeed the case, i.e. such a construction is possible around
any spacetime point $P$. It would also turn out that in this construction, the metric is of the same form as in the 
rest frame of any locally accelerating observer. 
To simplify the form of the metric, we are now going to make some further assumptions. First, we choose it to be of 
the form, $ds^2= g_{00}(dx^0)^2+dn^2+\sigma_{AB}dx^Adx^B$, where $A$ and $B$ can run from 1 to 2. We also 
demand that $t^{\alpha}_{;\beta}e^{\beta}_A= 0$ where $A={1,2}$. Two points should be noted here. Firstly, 
the direction corresponding to the vector $(0,dn,0,0)$ is being singled out. This is because the 
observer is accelerating along this direction. In other words, $\mathbf{dn}$ is chosen to be parallel to 
the acceleration vector $\mathbf{a}$. The second point is that we are also assuming that the velocity vector 
field does not change as we move along the other two spacelike directions within the local patch.

We shall now determine the form of the metric in the C.S. of $A$. This will also demonstrate that 
\eqref{Killingchange} can always be satisfied up to first order within the local patch that we have 
been considering. The acceleration vector $\mathbf{a}$ is given by
\begin{equation}
 t^{\alpha};_{\beta}t^{\beta} = a^{\alpha} . \label{accn}
\end{equation}
The norm of $\mathbf{\xi}$ is denoted as
\begin{equation}
 -\mathbf{\xi^{\mu}}\mathbf{\xi_{\mu}}= \kappa_{\xi}^2 = \kappa . \label{xinorm}
\end{equation}
Since $t^{\alpha}$ is parallel to the Killing vector $\mathbf{\xi^{\alpha}}$, we also have
\begin{equation}
 t^{\alpha}= \frac{\xi^{\alpha}}{\kappa_{\xi}}  .  \label{txi}
\end{equation}
Denoting the proper time along the trajectory of the accelerated observer by $\tau$, we can write 
\begin{equation}
 \frac{d\mathbf{\xi_{\mu}}}{d\tau}= \kappa_{\xi,\mu},   \label{xitau}
\end{equation}
where we have used the fact that $t^\mu$ is a unit vector and also the Killing equation for $\mathbf{\xi}$.
From the Killing equation, one also gets 
\begin{equation}
 \mathbf{\xi^{\mu}}_{;\nu}\mathbf{\xi^{\nu}} = \kappa_{\xi} \kappa_{\xi,\nu}  \label{a}
\end{equation}
and using \eqref{a}, 
\begin{equation}
 a= (\ln{\kappa_\xi})_{,\alpha}n^{\alpha}= \frac{\partial}{\partial n}(\ln{\kappa_\xi}) . \label{an}
\end{equation}
This fixes the $g_{00}$ component of the metric. To fix the other components, we note that $\xi_{\alpha;\beta}n^{\beta}
= a\xi_{\alpha}+\kappa_{\xi}t_{\alpha;\beta}n^{\beta}$. Then using \eqref{Killingchange}, we get
\begin{equation}
 t_{\alpha;\beta}n^{\beta}= 0 . \label{tn}
\end{equation}
This fixes the metric to be of the form,
\begin{equation}
 ds^2= -e^{2an} dt^2+ dn^2 + \sigma_{AB}(x^1,x^2)dx^Adx^B . \label{metric}
\end{equation}
Since $t^\beta$ and $n^\beta$ are orthogonal to each other, using the Killing equation, it follows that
\begin{equation}
 \mathbf{\xi}_{\alpha;\beta}n^{\beta} = n_{\beta;\alpha}\xi^{\beta} . \label{xin}
\end{equation}
From \eqref{xin} and \eqref{Killingchange}, one gets $a\xi_\alpha = n_{\beta;\alpha}\xi^{\beta}$. This equation 
implies 
\begin{equation}
 a g_{\alpha \beta} = n_{\beta;\alpha}+ f(x^\mu) h_{\alpha \beta}, \label{proje}
\end{equation}
where, $h_{\alpha \beta}= g_{\alpha \beta}+t_\alpha t_\beta$, is the projection operator on surfaces orthogonal 
to $t_\alpha$ and $f(x^\mu)$ is an 
arbitrary function of $x^\mu$. To fix $f$, we take the trace on both the sides of \eqref{proje} and obtain
\begin{equation}
 4a = k+3f ,     \label{excurv}
\end{equation}
where $k= n^\alpha_{;\alpha}$ is the extrinsic curvature scalar of the two dimensional hypersurface orthogonal to $n_\mu$. 
To determine $f$, we contract \eqref{proje} by $n^\mu$. This gives $f=a$. Putting it back in \eqref{excurv}, we have
\begin{equation}
 k=a .     \label{ka}
\end{equation}
It is interesting to note that the acceleration and hence also the temperature seen by the accelerating observer 
comes out to be the extrinsic curvature scalar of the two dimensional spacelike hypersurface orthogonal to the 
direction of acceleration. Another point should also be mentioned here. Near the event horizon of a black hole 
and also near the Rindler horizon, the null vector normal to the horizon and the stationary Killing vector obeys a relation 
similar to the one between $n_\mu$ and the timelike Killing vector $\xi^\mu$, as given by \eqref{Killingchange}. 
However, at the horizon, the stationary Killing vector itself becomes a null vector normal to the horizon. 
\let\thefootnote\relax\footnotetext{The Killing vector field vanishes at the bifurcation surface for Rindler spacetime.}

Now, we shall show that given a spacetime, locally it is always possible to construct such a C.S.. 
For this, let us first consider 
the null geodesics in the $t-n$ plane. They are given by 
\begin{equation}
 t= \mp \frac{1}{a}(e^{-an}-1)  .  \label{gd} 
\end{equation}

We define the null coordinates $u$ and $v$ by 
\begin{equation}
 u = -t+\frac{1}{a}(e^{-an}-1); v= t+\frac{1}{a}(e^{-an}-1) .         \label{uv}
\end{equation}
Denoting the $t-n$ part of the metric by $ds_{1+1}^2$, we have 
\begin{equation}
 ds_{1+1}^2= e^{2an}dudv . \label{metricuv}
\end{equation}
Now we are going to use the condition that the metric is being determined only in a local patch.
Using the fact that $t\ll1$ and $n\ll1$ and ignoring the second order terms, one gets $u\approx -t-n$ and $v\approx t-n$. 
From the new expressions of $u$ and $v$, it is possible to write $2n\approx -(u+v)$. We also note that within 
this scheme of approximation, \eqref{metric} can be expressed as 
\begin{equation}
 ds^2 \approx -(1+2an)dt^2 + dn^2+ + \sigma_{AB}(x^1,x^2)dx^Adx^B .
\end{equation}
In what follows, we shall freely 
interchange the exponential functions of the variables $u$,$v$,$t$ and $n$ with their polynomial forms approximated up to 
the appropriate order in these variables. Such manipulations do not lead to any inconsistency, as long as we keep in mind 
that they represent the functions (like exponentials) only up to a given order in these variables.

One can then express \eqref{metricuv} as $ds_{1+1}^2\approx e^{-2a(u+v)}dudv$. This can be rewritten as 
 \begin{equation}
 ds_{1+1}^2\approx d\bar{u}d\bar{v}, \label{Min}
\end{equation}
where 
\begin{equation}
 \bar{u}\approx \frac{e^{-au}}{a};  \bar{v}\approx \frac{e^{-av}}{a}.  \label{exp}
\end{equation}
This brings the $t-n$ part of the metric to a Minkowskian form. As there is no constraint on the $\sigma_{AB}(x^1,x^2)$
part of the metric, it can always be brought to an Euclidean form. 
It should be noted here that in the above analysis, we have 
always made the approximations in such a way, so that the metric contains terms only up to first order. 
\eqref{exp} gives the relation between the two sets of null coordinates locally, and they contain terms up to 
second order. \eqref{Min} establishes 
two things. Firstly, we note that up to the first order, locally it is always possible to transform from the C.S. in the $A$ frame defined
 by us to one in which the metric is Minkowskian and vice versa. As it is always possible to express the metric locally(i.e. up to the first 
order) in a Minkowskian form, this demonstrates that the assumptions made by us about the properties of the Killing vectors defined locally 
and so on are justified. The second point is that the relation between the null coordinates in the Minkowskian frame and the 
accelerating frame $A$ can be expressed in a form similar to that between the Rindler C.S. and Minkowskian one up to second order 
in $u$ and $v$.
% This means that the relation between the two sets of null coordinates is the same up to second order terms.
%{\bf ( As we shall see, this fact is going to be important later on.)}

\subsection{Heat generation induced by Gravity and the derivation of Einstein equation}

We are now in a position to proceed to the thermodynamic analysis. The basic idea is very simple. For a 
causal diamond constructed locally in the way described earlier, the increase in gravitational entropy 
is assumed to be proportional to 
the amount of increase in the area  of the cross-section of the null geodesic congruences along the properly 
specified part of the null boundaries of the causal diamond.  A certain fraction 
of the energy flux crossing this diamond is assumed to be converted into heat. The amount of heat generated and the 
increase of the entropy are then related via the relation, \dj{}$Q= Tds$. Plugging in the expressions for these quantities, 
the Einstein equation is expected to follow.

The most important point on which, the construction given here differs from that in Jacobson's paper 
\cite{Jacobson}, is in determining the amount of heat generated. In Jacobson's  construction, the Rindler 
horizon constitutes a part of the boundary of the locally constructed causal patch. The energy flux passing 
through the horizon is then taken to be the amount of energy that gets transformed into heat. The 
justification for this choice comes from many facts. The fact that thermodynamic laws are satisfied by the 
Rindler-like horizon itself provides 
strong support for this \cite{{JacobParent}}. Also, within the membrane paradigm, one finds further evidence 
in support of such a view.
The horizon acts like an absolute screen for the accelerating observer, who cannot observe the degrees of freedom outside the 
horizon. So it makes sense to interpret the energy crossing the horizon as heat energy. 
But for the causal diamond that we are considering, the null boundaries do not constitute an absolute screen in the sense a horizon 
does. It only demarcates an effective boundary of a local patch, where one can define a stationary Killing vector up to first order
accuracy. 
So we need a new prescription or method to determine the increase in heat energy. Moreover, such a prescription 
should also be able to do two things. Firstly, it should be able to reproduce the Einstein equation. It should also 
be able to include the fact that the Rindler-like observer sees the energy passing through the horizon as heat.
Such a prescription is proposed below.\\

{\it In a locally accelerating frame, there exists a heat flux due to gravity which heats up the null surfaces 
within the local patch. Let, $\delta\mathbf{\xi}$ be the change in the stationary Killing vector $\mathbf{\xi}$, due 
to the non-inertial motion of the observer as one moves orthogonally across any null hypersurface within this 
patch. Then $\delta\mathbf{\xi}$ gives rise to a heat flux given by $T_{\mu\nu}\delta\xi^\mu \mathbf{k^\nu}$. The total amount of heat 
generated in this way in a locally accelerating frame, can be computed using this prescription.}

It is easily seen that this prescription gives the desired expression for the increase in heat energy in the 
case considered by Jacobson \cite{Jacobson}. In that case, the amount of matter-energy crossing the horizon, 
that is converted into heat in the frame of the Rindler observer, can be determined by using it. The coordinate 
system of the accelerating observer, which is locally Rindler, breaks down at the
causal boundary , i.e. the Rindler horizon. Hence, the change in the stationary Killing vector or the Boost Killing vector
as one crosses the horizon is taken to be equal to $\mathbf{\xi}$ itself. This gives the standard expression 
for the increase in heat in this case. 
% It should be emphasized here that the local patch for a Rindler-like observer is very thin along $n$ on the 
% $n,t$ plane and for all practical purposes, the change in the Killing vector 
% $\mathbf{\xi}$ can be taken to be its change in crossing the horizon. (Probably another way to understand 
% this is to consider the very high blue shift factor that comes from near the horizon and which falls off sharply as one moves away from it. This implies that 
% $T_{\mu\nu}\delta\xi^\mu \mathbf{k^\nu}$ would have a value much higher near the horizon than further away from it.)
% {\bf( It should also be noted that the Killing vector in question is timelike for the Rindler-like observer, 
% though very close to a null vector near the horizon. It is a (degenrate?) null vector only at the horizon. 
% This implies that the Killing vector in this case can be considered a stationary killling vector inside the 
% local patch.[Should be put in a footnote])}

With this, we are in a position to derive the Einstein equation. First, let us focus on the amount of energy that
would be converted into heat in the locally accelerated frame according to the prescription outlined above. 
Let, $k^\mu= \frac{dx^\mu}{d\lambda}$ be the 
tangent vector to a null geodesic. Since both $n^\mu$ and $n_\mu$ are constant vector fields within $\mathscr{D}$, 
one can write 
\begin{equation}
 d\lambda= \frac{dn}{k^\mu n_\mu},   \label{lambdan}
\end{equation}
where $dn$ is the amount by which $n$ changes along the null geodesic. This formula would be useful while 
performing integrations along null geodesics.

Now according to the above discussion, the heat flux is to be computed from the change 
in $\mathbf{\xi}$ as one moves across a null hypersurface orthogonally and also it is the null hypersurfaces 
that heat up in this process. So it would be convenient if we first decide on what are the null hypersurfaces 
that we have to take into account to determine the total heat increase as seen by the observer. From the 
form of the metric \eqref{metric}, it is clear that the Killing vector $\mathbf{\xi}$ does not change 
as we move within the $(t,x^1,x^2)$ subspace. Hence we need to consider only the null hypersurfaces within the 
$(t,n)$ subspace. These are a series of $u=const$ and $v=const$ hypersurfaces, which crisscross each other 
within the causal diamond. Now, within $\mathscr{D}$, there is a change in $\mathbf{\xi}$ as we move across 
one such $u=const$ surface and also across one such $v=const$ surface. This means that we have to 
add the heat fluxes due to these two sets of null surfaces. Since each null hypersurface gets heated up 
by the resultant heat flux, we have to compute the total amount of increase in heat by multiplying this 
heat flux with the appropriate volume form for each set of null surfaces and then adding them up.

To compute the amount of heat increase, we can start from the null surface $CD$, which is a surface on which 
$u=const$. Now starting from this null surface, we want to move across other $u=const$ surfaces towards 
the center of $\mathscr{D}$, i.e. in the direction $n$ decreases. For this purpose, we choose the null vector 
$\mathbf{k_u}$ normal to $u=const$ hypersurfaces, where $k_u^\mu\approx -t^\mu-n^\mu$. The part of 
the heat flux generated due to the change of $\mathbf{\xi}$ across $u=const$ surfaces is given by 
$T_{\mu\nu}\delta_{\mathbf{k_u}}\xi^\mu$, where, $\delta_{\mathbf{k_u}}\mathbf{\xi}^\mu= \mathbf{\xi}^\mu;_\nu k_u^\nu \delta \lambda_u $. 
The total heat energy flux is given by, $h^\mu= T_{\mu\nu}\delta_{\mathbf{k_u}}\xi^\mu+T_{\mu\nu}\delta_{\mathbf{k_v}}\xi^\mu$, 
where, $\mathbf{k_v}$ is the appropriate null vector normal to $v=const$ surfaces and 
$\delta_{\mathbf{k_v}}\mathbf{\xi}^\mu= \mathbf{\xi}^\mu;_\nu k_v^\nu \delta \lambda_v $. $\lambda_u$ and $\lambda_v$ 
are the affine parameters along the null geodesics, the tangent vectors to which are $\mathbf{k_u}$ and $\mathbf{k_v}$ 
respectively. To choose $\mathbf{k_v}$, we start from the null boundary $DA$ and again move inwards from it. 
This fixes $k_v^\mu\approx t^\mu-n^\mu$. Any $u= const$ surface gets heated up due to the sum of the heat 
fluxes which are proportional to 
$\mathbf{\delta_{k_u}\xi}$ and $\mathbf{\delta_{k_v}\xi}$. The $v=const$ null surfaces heat up in the same way. 
One can find out the total amount of increase in heat energy by adding up the contributions of all these four parts.

They are given by\\
$dQ_{uu}= \int T_{\mu\nu}(\delta_{\mathbf{k_u}}\mathbf{\xi}^\mu) k_u^\nu d\lambda_u dA$, \\
$dQ_{vu}= \int T_{\mu\nu}(\delta_{\mathbf{k_v}}\mathbf{\xi}^\mu) k_u^\nu d\lambda_u dA$, \\
$dQ_{uv}= \int T_{\mu\nu}(\delta_{\mathbf{k_u}}\mathbf{\xi}^\mu) k_v^\nu d\lambda_v dA$, \\
$dQ_{vv}= \int T_{\mu\nu}(\delta_{\mathbf{k_v}}\mathbf{\xi}^\mu) k_v^\nu d\lambda_v dA$, \\
where, $\delta_{\mathbf{k_u}}\mathbf{\xi}^\mu= \mathbf{\xi}^\mu;_\nu k_u^\nu \delta \lambda_u $ and $\delta_{\mathbf{k_v}}\mathbf{\xi}^\mu= \mathbf{\xi}^\mu;_\nu k_v^\nu \delta \lambda_u$ and 
$\lambda_u$ and $\lambda_v$ are the affine parameters along the null geodesics, the tangent vectors to which are $\mathbf{k_u}$ and $\mathbf{k_v}$ 
respectively. Also $dA$ is the area element of the cross section of the null hypersurfaces of the causal diamond. The volume element 
$\mathbf{k}d\lambda dA$ is being used as we want to determine the total heat energy transferred to the null surfaces 
orthogonal to $\mathbf{k}$
within the causal diamond. From \eqref{lambdan}, we have $d\lambda_u = \frac{dn}{k_u^\mu n_\mu}$ and $d\lambda_v= \frac{dn}{k_v^\mu n_\mu}$ and also 
similar relations for $\delta \lambda_u$ and $\delta \lambda_v$. Using them, the total amount of increase in 
heat energy can be expressed as  
\begin{equation}
 \dbar Q \approx 4 \int T_{\mu\nu}(\delta_{\mathbf{n}}\mathbf{\xi}^\mu) n^\nu dn dA ,
\end{equation}
where the relations $k_u^\mu n _\mu \approx -1$ and $k_v^\mu n _\mu \approx -1$ have also been used. Using \eqref{Killingchange}, and  
using $n$ and $\delta n$ interchangeably(since $n\ll1$), one can express the amount of heat increase as 
\begin{equation}
 \dbar Q \approx 4a \int_0^n T_{\mu\nu} \mathbf{\xi}^\mu n^\nu  n dn dA . \label{heatgn}
\end{equation}
One important point to note here is that the amount of the increase in heat energy depends upon the fact 
that it is the null surfaces that are being heated up. This is a characteristic property of the gravitational 
heating process that is being considered here. As seen here, it has physical consequences.

Now we consider the right hand side of the equation, $\dbar Q= Tds$. The temperature $T$ is taken to be proportional to the acceleration $a$.
The justification for this is provided in the next section. Assuming the gravitational entropy to be holographic and taking a cue from black hole thermodynamics, the 
increase in entropy $ds$  is taken to be proportional to the increase of the area of the cross section of the null boundary of the diamond. This null boundary can be thought of 
as being constituted of two wedges and we consider the area increase along one of the wedges. Our calculation of the 
increase of the area of the cross section closely follows the one performed by Jacobson \cite{Jacobson}. However, there are some important 
points, on which our method differs from his.  These would be pointed out as we go along.

The increase of the area of the cross-section of the null boundary of $\mathscr{D}$ is calculated along the wedge $CDA$. 
We choose two null congruences along $CD$ and $DA$. They are chosen in such a way that for both the congruences, 
$\theta$ and $\sigma$ vanish at $C$. Now matter-energy passes through both the sides $CD$ and $DA$ of the null wedge. 
Hence the focusing or the defocussing effect of this matter on the null congruences would be such that along $CD$, the 
area increases and the expansion 
becomes zero at $C$. However, $\theta$ is negative along $DA$ and hence the area of the congruence decreases along it. 
%It is obvious that the congruences along $AB$ and $BC$ are different. 
We assume that $ds\propto (\delta A_{CD}+ \delta A_{DA})$, where $\delta A_{CD}$ 
and $\delta A_{DA}$ are the changes in the area of the null congruences along $CD$ and $DA$ respectively. Now one potential problem with 
this assumption is that $ds$ might turn out to be negative. However, as we shall see, in this case, it turns out that although 
$\delta A_{CD}>0$ and $\delta A_{DA}<0$, $(\delta A_{CD}+\delta A_{DA} )>0$. So at least this problem of decrease in entropy 
is avoided. The fact that the sign 
of the change in the area of the null congruences is opposite along the sides $CD$ and $DA$ implies that the change in this area along the
boundary of the causal diamond does not increase monotonically with time in the locally accelerating frame. Instead, it increases first 
till the time slices evolve up to the hypersurface $BD$ and then start decreasing. This means that even though the entropy is proportional
to the total change in the area of the null congruences along the null boundary, within the causal diamond itself, one cannot assign it 
to be proportional to the incremental change in area with time. In other words, using this prescription, one can only
determine the total entropy increase in the causal diamond $\mathscr{D}$ as a whole. This feature of the entropy as defined 
here contrasts strikingly with that of the 
horizon entropy, which increases monotonically, once a particular type of of teleological boundary condition 
has been assumed. Using this prescription, one can only find out about the overall increase in entropy within 
the locally constructed causal diamond. We shall come back to this issue again in the final section of this paper.

The task of calculating the change in the area of the boundary, $\mathcal{B}$, in the way outlined above is straightforward. The change in the area is 
\begin{equation}
 \delta A = \int_{\mathcal{B}} \theta d\lambda dA,   \label{areainc}
\end{equation}
where, $\lambda$ is the affine parameter along the null geodesic congruence. Now the Raychaudhury equation for the null congruence 
is given by, 
\begin{equation}
 \frac{d\theta}{d\lambda} = -\frac{1}{2} \theta^2 - \sigma^{\mu\nu}\sigma_{\mu\nu} - R_{\mu\nu}k^\mu k^\nu,  \label{RCE}
\end{equation}
where, $\sigma_{\mu\nu}$ denotes the shear of the congruence and $\mathbf{k}$ is a null vector tangent to the null geodesics.
We choose the boundary in such a way that $\theta$ and $\sigma$ vanishes
at $D$. Neglecting the higher order terms, one gets from \eqref{RCE}, $\theta=  -\lambda R_{\mu\nu}k^\mu k^\nu$. Putting this back in 
\eqref{areainc}, we have 
\begin{equation}
 \delta A= - \int_{\mathcal{B}} \lambda R_{\mu\nu}k^\mu k^\nu d\lambda dA . \label{genarea}
\end{equation}
The null vectors tangent to the sides $CD$ and $DA$ that we are going to use below are denoted by $\mathbf{\bar{k}_u}$ 
and $\mathbf{\bar{k}_v}$ respectively. They are given by the relations $\bar{k}^\mu_u \approx t^\mu+n^\mu$ and 
$\bar{k}^\mu_v \approx t^\mu-n^\mu$. The total area change along the boundary can be written as
\begin{equation}
 \delta A =  -(\int_{CD}\lambda_u R_{\mu\nu}\bar{k}_u^\mu \bar{k}_u^\nu d\lambda_u dA + \int_{DA}\lambda_v R_{\mu\nu}\bar{k}_v^\mu \bar{k}_u^\nu d\lambda_v dA) . \label{arinc1}
\end{equation}
It is important to recall here that $\lambda_u$ increases monotonically along $CD$ from a negative value at $C$ to zero value at $D$, 
whereas $\lambda_v$ increases monotonically along $DA$ from zero value at $D$ to a positive value at $A$. Let us assume that 
$n$ increases along $CD$ from zero value at $C$ to $n_{max}$ at $D$ and then again decreases along $DA$, becoming zero 
at $A$. Then changing the variables from $\lambda_u$ and $\lambda_v$ to $n$, we get 
\begin{equation}
 \delta A= - \int^{n_{max}}_0(n-n_{max})R_{\mu\nu}(\bar{k}^\mu_u\bar{k}^\nu_u- \bar{k}^\mu_v\bar{k}^\nu_v) dn dA .\label{eq1}
\end{equation}
Since $R_{\mu\nu}$ is a symmetric tensor, one may add a term $ R_{\mu\nu}(k_u^\mu k_v^\nu-k_v^\mu k_u^\nu)$ in the integrand of the 
R.H.S. of \eqref{arinc1},  without changing anything. Then defining a variable, $x= n-n_{max}$, we get 
\begin{equation}
 \delta A = 4 \int_0^{-n_{max}} R_{\mu\nu} t^\mu n^\nu x dx dA. \label{arinc2}
\end{equation}
Now ignoring second order terms in $x$ in the integrand of \eqref{arinc2}, 
\begin{equation}
 \delta A = 4 \int_0^{n_{max}} R_{\mu\nu} t^\mu n^\nu n dn . \label{arinc3}
\end{equation}
The upper limit of $n$ in the $n$-integral of \eqref{heatgn} is also $n_{max}$.
Now assuming that $\dbar Q= Tds$ holds, and using \eqref{heatgn} and \eqref{arinc3}, one gets 
\begin{equation}
 T_{\mu\nu} \mathbf{\xi}^\mu n^\nu = R_{\mu\nu} t^\mu n^\nu. \label{genE}
\end{equation}
\eqref{genE} can be written as $\kappa_\xi T_{\mu\nu} t^\mu n^\nu = R_{\mu\nu} t^\mu n^\nu$. Now neglecting the first order terms, one 
gets 
\begin{equation}
 T_{\mu\nu} t^\mu n^\nu = R_{\mu\nu} t^\mu n^\nu . \label{genE2}
\end{equation}
Since $t^\mu$ and $n^\mu$ are orthogonal and one can choose $t^\mu$ as one wishes, the following identity holds true,
\begin{equation}
 T_{\mu\nu}+f(x^\alpha)g_{\mu\nu} = R_{\mu\nu}  \label{genE3}
\end{equation}
It can be easily shown that if the energy-momentum tensor is assumed to be conserved, then it follows that, 
$f=  -\frac{R}{2} +\Lambda$, where $\Lambda$ is a constant and one ends up with the Einstein equation along with a cosmological 
constant term.

\section{ Detection of Thermal Environment by a Locally Accelerating Observer}
In this section, it is shown how, an observer at rest in a locally accelerating frame, would encounter a thermal environment. Before 
proceeding 
with the analysis, the notion of the locally accelerating observer has to be made more precise. Specifically, the key feature 
distinguishing any such observer is that the scales at which she could make a measurement is much smaller than the size 
of the local patch. It would also be assumed that the physics at this smaller scale would be local up to a very good 
approximation, i.e. it would not be influenced by physics at a scale, which is much larger than the size of the local patch. For the  
patch that we are considering, it is also possible to transform from the C.S. in which the accelerating observer is at rest to the C.S. in 
which the inertial observer is at rest. Since local observers can observe physics at a much smaller scale than the size of this 
patch, it can be treated as the Minkowski spacetime as far as they are concerned. These considerations should hold in both these 
frames up to a very good approximation. 
%Formally, this is done by taking the limit $n \rightarrow \infty$, $t \rightarrow \infty$ and $x^{[1,2]} \rightarrow \infty$. 

Now, it would be further assumed that within these approximations, the 'inertial' observer is going to see 
a state analogous to a vacuum state in the Minkowski spacetime in the inertial 
frame. So one has to construct such states. This can be done as follows. First, one focuses on a particular wave 
equation of a free field, like the wave equation for 
the massless Scalar field. Then one tries to build Fock states by taking into account only the short wavelength modes associated with 
the local patch or in other words, by excluding the wave modes which are of wavelength larger than the size of the local patch. In 
this approximate sense, one can talk about the different Fock states associated with this local patch. The assumption made here is that 
that the inertial observer associated with this local patch would not detect any particles, i.e. the 
inertial observer finds the field to be in the vacuum state. In a similar way, it is also possible to construct 
the Fock states using the short wavelength modes associated with the local patch for the observers, who are 
accelerating. What will be shown here is that in this case, those accelerating observers find themselves in a 
thermal environment.

To deal with the wave modes, which are much smaller than the size of the local patch that we have been considering, let us define 
the quantities, $t_s= \Lambda t$, $n_s= \Lambda n$, $x^1_s= \Lambda x^1$ and $x^2_s= \Lambda x^2$. These 
are just the rescaled time and space coordinates. The rescaling factor is a constant $\Lambda$, chosen in such a way, that 
$\Lambda L\gg 1$, where $L$ is the size of the patch. This can be thought of as introducing very small units of length, 
which the observer is using to measure length and time,
 i.e. the length resolution power of the local observers is much higher. The value of any quantity after such a rescaling
of the coordinates is going to be 
denoted here by the suffix 's'. The value of the acceleration after such a rescaling becomes 
$a_s= \frac{a}{\Lambda}$. 
This means that in the rescaled units, $a_s\ll1$. Apart from being multiplied by a constant factor, the 
metric retains the same form in the accelerated frame after the rescaling of the coordinates. We consider 
only the $(t,n)$ part of the metric here. Apart from a constant factor, it can be written as 
\begin{equation}
 ds^2_{1+1}= -e^{2a_sn_s}dt_s^2 + dn^2 . \label{metrics}
\end{equation}
One can now define $u_s$ and $v_s$, which are of exactly the same form as in \eqref{uv} with only $a_s$, $t_s$ and $n_s$ replacing $a$, 
$t$ and $n$ respectively. Consequently, \eqref{metrics} can be written as $ds^2_{1+1}= e^{2a_sn_s} du_s dv_s$ in the same way as 
\eqref{metricuv} was deduced. It is important to note here that $a_sn_s= an$. Hence $a_sn_s\ll1$. In the same way as in \eqref{exp}, 
one can define $\bar{u}_s$ and $\bar{v}_s$ by the relations 
\begin{equation}
 \bar{u}_s\approx \frac{e^{-a_su_s}}{a};  \bar{v}_s\approx \frac{e^{-a_sv_s}}{a}.  \label{exps}
\end{equation}
One can transform to the coordinates $\bar{u}_s$, $\bar{v}_s$ and write the metric in a Minkowskian form,
\begin{equation}
 ds_{1+1}^2\approx d\bar{u}_sd\bar{v}_s.  \label{Mins}
\end{equation}

With the coordinates rescaled, we now move on to the issue of constructing the Fock states for the local observers in the inertial and 
in the accelerating frames. For this purpose, we shall consider a free massless scalar field $\phi$ here. Within the scheme of 
approximation outlined above, one can expand $\phi(t_s,\vec{x_s})$ in terms of the orthonormal modes as 
\begin{equation}
\phi(t_s,\vec{x_s}) = \sum\limits_{\mathbf{k}}[a_{\mathbf{k}}\phi^{\bar{u}}_{\mathbf{k}}(t_s,\vec{x_s})+ a_{\mathbf{k}}^{\dagger}\phi^{\bar{u}}_{\mathbf{k}}(t_s,\vec{x_s})^*] . \label{modeexpin}
\end{equation}
In a similar way, in the accelerated frame also, $\phi(t_s,\vec{x_s})$ can be expanded in terms of the orthonormal modes denoted by 
$\phi^u_{\mathbf{k}}(t_s,\vec{x_s})$ as 
\begin{equation}
 \phi(t_s,\vec{x_s}) = \sum\limits_{\mathbf{k}}[a_{\mathbf{k}}\phi^u_{\mathbf{k}}(t_s,\vec{x_s})+ a_{\mathbf{k}}^{\dagger}\phi^u_{\mathbf{k}}(t_s,\vec{x_s})^*] . \label{modeexpaccn}
\end{equation}
The modes denoted by $\phi^{\bar{u}}$ are just the plane wave modes in the Minkowskian spacetime. However one needs to 
look at the modes denoted by $\phi^u$ in the accelerated frame more carefully. The wave equation for the massless scalar 
field $\phi$ can be written down in the accelerated frame as 
\begin{equation}
 -\partial_{t_s}^2 \phi - a_s \partial_{n_s}\phi \partial_{t_s}\phi + \partial_{n_s}^2\phi+ \sum\limits_{i=1}^2\partial_i^2\phi = 0 . \label{weaccn}
\end{equation}
At this point, it is important to realise that the term proportional to $a_s$ in the L.H.S. of \eqref{weaccn}
is going to be much smaller in magnitude than the other terms, since $a_s\ll1$. Then one can treat this term 
as a perturbation to the free scalar field wave equation in Minkowski spacetime in the inertial frame. If the 
wave equation is then solved as a perturbation series order by order, then at the zeroeth order, \eqref{weaccn}
reduces to the standard wave equation in the inertial frame. The general solution to it can be written as a 
Fourier series of the plane wave modes. Even when one solves the equation in the 
first order, one can express the general solution as a Fourier series in terms of the same plane wave modes. 
Up to first order then, these plane wave modes constitute a complete orthonormal basis for \eqref{weaccn}. 
This fact is going to be used in the analysis that follows. That these plane wave modes from an orthonormal 
set is a well known fact and will be assumed as such.

Alternatively, this also follows from the fact that \eqref{weaccn} can be expressed in the form 
\begin{equation}
 \frac{\partial^2\phi}{\partial_{u_s}\partial_{v_s}}= 0, \label{weaccnull}
\end{equation}
in terms of the null coordinates $u_s$ and $v_s$. An equation of the form 
\eqref{weaccnull} has plane wave modes as solutions [See \cite{Birell}].

It should be emphasised here that it is only within a scheme of approximation, that these wave modes 
could be taken to form an eigenbasis in both the frames. The size of the patch being considered 
is of the order $\Lambda L$. So one can solve the massless Klein-Gordon equation with periodic boundary 
conditions in a box of size $\Lambda L$. Then the wavenumbers are discrete and given by 
$\frac{n\pi}{\Lambda L}$, where $n$ belongs to the set of all positive integers. Now instead of taking the 
limit of $\Lambda L$ going to infinity, we just consider the case when $\Lambda L$ is very large. 
Then we can replace the discrete wavemodes by the continuous ones up to a very good approximation. 
It is only within this approximation that the operators $a$ and $a^\dagger$ are to be defined and the commutation 
relation between them would hold. In fact, in the commutation relation between $a$ and $a^\dagger$, 
the right hand side can be thought of as a properly weighted sum of the discrete wave modes, which is a good approximation to
a Dirac delta function, when $\Lambda L$ is very large. However, one can also scale the $x^\mu$s to get 
rid of the factor $L$ and 
absorb it within $a$ and $a^\dagger$ as it is a constant factor. After that, only $\Lambda$, which is a 
very large 
number, remains in the denominator of the wave number. Then arguing in the same way as before,   
one can approximate this sum as a Dirac delta function.

Once the basis sets have been constructed in both the frames, we can ask the question, {\it 'What state does 
the vacuum state in the inertial frame correspond to in the accelerated frame?'}. This can be answered by 
calculating the Bogulobov coefficients pertaining to this case. We shall now calculate these coefficients 
by closely following a similar calculation done by Visser {\it et al}\cite{Visser}. A similar approach has 
also been followed in \cite{VisserPRD}. The Bolgulobov coefficients are given by 
\begin{eqnarray}
 \alpha(\omega, \bar{\omega})&=& \left (\phi^u(\omega;t,\vec{x}), \phi^{\bar{u}}(\bar{\omega};t,\vec{x}))\right. \nonumber\\
                             &=& \left.-i\int d^3x \{\phi^u(\omega;t,\vec{x})\partial_t\phi^{\bar{u}}(\bar{\omega};t,\vec{x})^*
                             -\phi^{\bar{u}}(\bar{\omega};t,\vec{x})^*\partial_t\phi^u(\omega;t,\vec{x})\}\right.  \label{Bogualpha}
\end{eqnarray}

and 

 \begin{eqnarray}
  \beta(\omega, \bar{\omega})&=& \left. -(\phi^u(\omega;t,\vec{x}), \phi^{\bar{u}}(\bar{\omega};t,\vec{x})^*)\right. \nonumber\\
                              &=& \left. -i\int d^3x \{\phi^{\bar{u}}(\bar{\omega};t,\vec{x})\partial_t\phi^u(\omega;t,\vec{x})
                              -\phi^{u}(\omega;t,\vec{x})\partial_t\phi^{\bar{u}}(\bar{\omega};t,\vec{x})\}\right. . \label{Bogubeta}
 \end{eqnarray}

The integrals are over any arbitrary spacelike hypersurface within the local patch and extends at least 
till the boundary of the local patch. This can be thought of as the analogue of a spacelike hypersurface 
terminating at the spacelike infinity in a Minkowskian spacetime. In fact, as we have argued earlier, once we
rescale the coordinates,
the local patch can be treated as Minkowskian up to a very good approximation. This implies that in our 
case also, the Bogulobov coefficients are Lorentz invariant and hence do not depend on the hypersurface on which
they are evaluated. The frequencies $\omega$ and $\bar{\omega}$ correspond to the frequency of the wave modes
in the accelerated frame and in the inertial frame respectively. In this case, it is convenient to evaluate 
them on a spacelike hypersurface arbitrarily close to the future null boundary of the local patch that we 
have been considering. This means that we have to integrate over both a $u= constant$ and a $v= constant$ 
surface. Since we have chosen the $\phi^u$ modes as the basis, the integrand vanishes on the $v= constant$ 
surface. Hence, for the coefficient $\beta$, we get 
\begin{equation}
 \beta(\omega, \bar{\omega}) \propto \frac{1}{\sqrt{\omega \bar{\omega}}}\int_{u_{s-}}^{u_{s+}} du_s \{ exp[-i\omega u_s]\partial_{u_s} exp[-i\bar{\omega}\bar{u}_s]
                                     - exp[-i\bar{\omega}\bar{u}_s]\partial_{u_s} exp[-i\omega u_s]\},  \label{beta1}
\end{equation}
where, $u_s\in \{u_{s-}, u_{s+}\}$ is the range of $u_s$. $u_{s-}$ and $u_{s+}$ are both of the order of $\Lambda L$.

The first term can be integrated by parts giving rise to a surface term proportional to 
$exp[-i\omega u_s]exp[-i\bar{\omega}\bar{u}_s]\textbar^{u_{s+}}_{u_{s-}}$, which would be discarded. Thus 
one obtains the relation 
\begin{equation}
 \beta(\omega, \bar{\omega}) \propto \sqrt{\frac{\omega}{\bar{\omega}}}\int^{u_{s+}}_{u_{s-}}du_s \{exp[-i\omega u_s]exp[-i\bar{\omega}\bar{u}_s]\}. \label{beta2}
\end{equation}

Using the relation between $\bar{u}_s$ and $u_s$, \eqref{beta2} can be written as 
\begin{equation}
 \beta(\omega, \bar{\omega}) \propto \sqrt{\frac{\omega}{\bar{\omega}}}\int^{u_{s+}}_{u_{s-}}du_s \exp\large{[}-i\omega u_s+ i \bar{\omega}\frac{e^{-a_su_s}}{a_s}\large{]} . \label{beta3}
\end{equation}
It is important to keep in mind that in \eqref{beta3}, we have made use of the relation \eqref{exps} between $\bar{u}_s$ 
and $u_s$, where the exponential function is used only as an approximation. This can also 
be seen from the constraint, $a_su_s\ll 1$, that has to be satisfied. This constraint implies that even 
though $u_s$ becomes very large, $e^{-a_su_s}$ would always have a value close to $1$.

Now changing the variable from $u_s$ to $z$, where 
\begin{equation}
 z= \exp(-a_su_s),  \label{z}
\end{equation}
\eqref{beta3} can be expressed as 
\begin{equation}
 \beta(u_s;\omega, \bar{\omega}) \propto \sqrt{\frac{\omega}{\bar{\omega}}}\frac{1}{a_s} \int_{z(u_{s-})}^{z(u_{s+})} dz z^{i\omega/a_s-1}\exp(i\bar{\omega}\frac{z}{a_s}) . \label{zinteg}
\end{equation}
As has just been mentioned, the constraint forces both $z(u_{s-})$ and 
$z(u_{s+})$ to be close to $1$. Let us define a variable, $z_1=z-1$. Then the exponent in the integrand can 
be expressed as $e^{i\bar{\omega}}\exp(i\bar{\omega}z_1)$. The first factor does not depend on $z_1$ and can 
be taken out of the integral. $z_1$ would range from $0$ to $+\epsilon$ where 
$\textbar \epsilon \textbar \ll1$. Now the factor $\frac{1}{a_s}$ in the exponent of the integrand is very 
large. This means that the frequency of the waveform in the integrand is very high. In other words, the 
waveform varies very rapidly and so one can extend the range of integration from 
$\{ 0, +\epsilon \}$ to $\{-1, +\infty \}$. Extending the range of the integration contributes very little to 
the integral and can be neglected. One can go back to express the integral again in terms of $z$ as the 
variable and the range of integration is now extended from $0$ to $+\infty$. This is a crucial step in this 
derivation of the thermal spectra. As we shall see, the possibility of the extension of the integration 
range allows us to express the integral as a complete gamma function. This leads to a 
thermal spectrum. The extension of the range of the integration was possible because 
$a_s\ll1$. Another point to remember is that to obtain the thermal spectra, we need not take the limit 
$ \Lambda \rightarrow \infty $. The approximation to a thermal spectra is good enough for large values of $\Lambda$. 
This means that the rescaling of the coordinates in the local patch is directly responsible for 
the appearance of the familiar thermal weight factor. From a more physical perspective, the thermal nature is
resulting from the fact that the observer sees only local physics, i.e. the wave modes that are being observed 
are localised within the patch we have been considering. \\
So then, \eqref{zinteg} can be expressed as 
\begin{equation}
 \beta(u_s;\omega, \bar{\omega}) \propto \sqrt{\frac{\omega}{\bar{\omega}}}\frac{1}{a_s}\int_0^{+\infty}dz z^{i\omega/a_s-1}\exp(i\bar{\omega}\frac{z}{a_s}) . \label{zinteg1}
\end{equation}
The integral in \eqref{zinteg1} can be calculated as shown in \cite{Visser}. After neglecting an irrelevant phase, 
one gets
\begin{equation}
 \beta(u_s;\omega, \bar{\omega}) \propto \sqrt{\frac{\omega}{\bar{\omega}}}\frac{1}{a_s}\exp[-\pi\omega/(2a_s)]\Gamma(i\omega/a_s) .\label{gamma}
\end{equation}
Using the formula for Gamma functions,
\begin{equation}
 \textbar \Gamma(ix)\textbar^2 = \frac{\pi}{x sinh(\pi x)},
\end{equation}
one gets 
\begin{equation}
 \textbar \beta(u_s;\omega, \bar{\omega})\textbar^2 \propto \frac{\omega}{\bar{\omega}} \frac{1}{a_s^2}\pi \frac{\exp(-\pi\omega/a_s)}{\omega/a_s \sinh(\pi\omega/a_s)}. \label{thermsalsp1}
\end{equation}
One can express this in the familiar form of Planck spectrum,
\begin{equation}
 \textbar \beta(u_s;\omega, \bar{\omega})\textbar^2 \propto \frac{1}{a_s\bar{\omega}} \frac{1}{\exp(2\pi\omega/a_s)-1}. \label{Plancksp} 
\end{equation}
$\alpha$ can also be determined in a similar way. \\
\eqref{Plancksp} gives a temperature 
\begin{equation}
 k_B T= \frac{a_s}{2\pi}  . \label{Rindlertemp}
\end{equation}

There is however one serious problem in our derivation of the Planck spectrum. As we see from 
\eqref{Rindlertemp}, the temperature of the thermal radiation is proportional to $a_s$. Working in units such 
that $k_B=1$, one gets the wavelength of the temperature of the order of $\frac{1}{a_s}$ or $\Lambda$. This 
is of the order of the size of the local patch, we have been considering. This means that the temperature 
of the radiation is not a good observable quantity that could be measured by the observer in the local patch. 
In fact, the local observer cannot detect whether the radiation is in thermal equilibrium or 
not.  Hence it is not possible for the observer while remaining inside the local patch, to find out whether 
the state he is seeing is a thermal state or not. 
However, if the observer focuses on the localised modes or the higher energy modes, then he would find that 
these follow a Maxwell-Boltzmann spectrum, which is what the Planck spectrum tends to for the modes with energy much higher than 
the energy scale given by the temperature. 
%{\bf(In a sense, these small wavelength modes behave more like particles.)} 
So if {\it only} those localised wave modes are being 
measured by the observer, then he finds himself in a thermal environment. A possible way 
this could occur would be to use a detector at rest in the accelerating frame, coupled weakly to the massless scalar 
field. In this case, one finds the detector in equilibrium with a thermal 
environment, where the energy levels follow the Maxwell-Boltzmann distribution.

The above statement can be demonstrated by giving an example. One can consider a detector coupled to 
the massless scalar field. For the sake of concreteness, let us follow \cite{Birell} and  assume that the 
interaction Lagrangian is given by $cm(\tau_s)\phi[x_s(\tau_s)]$, where $c$ is a small coupling constant and $m$ is 
the detector's monopole moment operator. $x^\mu_s(\tau_s)$ gives the worldline, along which the detector moves 
and $\tau_s$ is the proper time along that worldline. We have already found out the state in which the field $\phi$ is in.
Let us denote this state by $|T\rangle$. We assume further that initially the detector is in the ground state with energy $E_0$. 
After interacting with the field $\phi$, it would undergo a transition to an excited state with energy  $E>E_0$, while the 
field would make a transition to an excited state $|\psi\rangle$. For small enough values of $c$, the 
amplitude of this transition, $\mathscr{A}$, is given by first order perturbation theory up to a good approximation. 
\begin{equation}
 \mathscr{A} = ic\langle E, \psi|\int^\infty_{-\infty}m(\tau_s)\phi[x_s(\tau_s)]d\tau_s |E_0, T\rangle  \label{transamp1}
\end{equation}
The equation for the time evolution of $m(\tau_s)$ is given by 
\begin{equation}
 m((\tau_s) = e^{iH_0\tau_s}m(0) e^{-iH_0\tau_s},  
\end{equation}
where, $H_0|E\rangle = E|E\rangle$. Using this, \eqref{transamp1} can be factorised and expressed as 
\begin{equation}
 \mathscr{A} = ic \langle E|m(0)|E_0\rangle \int^\infty_{-\infty}e^{i(E-E_0)\tau_s}\langle\psi|\phi(x_s)|T\rangle d\tau_s. \label{transamp2}
\end{equation}
To find out whether the detector detects a thermal environment or not, one needs to calculate the transition 
probability, $P$ of the state $|E_0,T\rangle$ to be excited to states with all possible values of energy $E$ and $\psi$ .
This is given by 
\begin{equation}
 P= c^2 \sum_E |\langle E|m(0)|E\rangle|^2 \Xi(E-E_0), \label{transprob1}
\end{equation}
where, 
\begin{equation}
 \Xi(E) = \int_{-\infty}^\infty d\tau_s \int_{-\infty}^\infty d\tau_s'e^{-iE(\tau_s-\tau_s')}\langle T|\phi[x_s(\tau_s)]\phi[x_s(\tau_s')]|T\rangle. \label{detresponse}
\end{equation}
If now the detector is in equilibrium with the field $\phi$ and moves along some worldline $x_s(\tau_s)$, then for 
those worldlines, 
\begin{equation}
 \langle T|\phi[x_s(\tau_s)]\phi[x_s(\tau_s')]|T\rangle = g(\Delta\tau_s), \label{ttranslate}
\end{equation}
where, $\Delta\tau_s= \tau_s - \tau_s'$. In this case, the number of quanta absorbed by the detector per unit 
proper time $\tau_s$ is constant. So the transition probability per unit proper time is given by, 
\begin{equation}
 \frac{dP}{d\tau_s} = c^2 \sum_E|\langle E|m(0)|E_0\rangle|^2\int_{-\infty}^{\infty}d(\Delta\tau_s)e^{-i(E-E_0)\Delta\tau_s} \langle T|\phi[x_s(\tau_s)]\phi[x_s(\tau_s')]|T\rangle . \label{transprobrate}
\end{equation}
Following \cite{Birell}, we can write this as 
\begin{equation}
 \langle T|\phi[x_s(\tau_s)]\phi[x_s(\tau_s')]|T\rangle = \langle 0|\phi(x_s)\phi(x_s')|0\rangle + \int d^3k\frac{n_{\mathbf{k}}}{|\mathbf{k}|}u_{\mathbf{k}}(x_s)u^*_{\mathbf{k}}(x_s') + \int d^3k \frac{n_{\mathbf{k}}}{|\mathbf{k}|}u^*_{\mathbf{k}}(x_s)u_{\mathbf{k}}(x_s'). \label{ampexpand}
\end{equation}
The factor $\frac{n_{\mathbf{k}}}{|\mathbf{k}|}$ comes from evaluating the Bogulobov coefficients.
From \eqref{ampexpand}, the detector response function can be computed assuming that it is at rest in the accelerating frame 
and it turns out to be 
\begin{equation}
 \frac{\Xi(E)}{\mathcal{T}} = \frac{1}{(2\pi)^3}\int_{-\infty}^{\infty}d(\Delta\tau_s)e^{-iE\Delta\tau_s}\int \frac{d^3k}{2\omega}\exp(i\omega\Delta\tau_s)\frac{n_{\mathbf{k}}}{|\mathbf{k}|}, \label{responserate}
\end{equation}
where, $\mathcal{T}$ is the time period over which the detector is switched on. The R.H.S. of \eqref{responserate} can be 
computed and one gets 
\begin{equation}
 \frac{\Xi(E)}{\mathcal{T}} \propto n_E. \label{reponsef2}
\end{equation}
Here one uses the fact that for a massless scalar field, $\omega=k$.
This in turn implies that 
% \begin{equation}
%  \frac{dP}{d\tau_s} \propto c^2\sum_E(E-E_0)|\langle E|m(0)|E_0\rangle|^2n_E  \label{transprob2}
% \end{equation}
% Since $n_E$ is the occupation number for the energy levels of a thermal state, \eqref{transprob2} can also 
% be expressed as 
\begin{equation}
 \frac{dP}{d\tau_s} \propto c^2\sum_E|\langle E|m(0)|E_0\rangle|^2e^{-\frac{2\pi(E-E_0)}{a_s}}. \label{detectorPS}
\end{equation}
From the Maxwell-Boltzmann exponential factor in the expression for the transition probability, it can be 
concluded that the detector is in a thermal equilibrium with the surroundings at a temperature $\frac{a_s}{2\pi}$. Here we have taken 
$n_E \propto e^{-\frac{2\pi(E-E_0)}{a_s}}$ as according to the argument given earlier, the
energy levels with lower values of energy would not be accessible to the detector. Hence the Planck factor 
goes over to the Maxwell-Boltzmann factor. It is also important to point out that in obtaining \eqref{transprobrate},
the closure relation for a complete set of $\lvert E\rangle$ s has been used. However, an assumption is made there 
that the range of $E$ in that relation roughly satisfies a lower bound given by $\frac{1}{E} \sim \Lambda L$. This can be 
viewed as a manifestation of the fact that the detector is disentangled from outside the local patch. This requirement 
of course is in line with our demand that the local observer does not see the physics outside of this small patch.

The factor $\exp[-\beta_s E]$ is taken here as the signature of the thermal environment. However, $T_s$ is very small. 
This makes $\beta_s$ very large. So we have to check whether the weight factor $\exp[-\beta_s E]$ associated with an 
energy level $E$ is going to be vanishingly small for all the excited states physically relevant for the locally 
accelerating observer.  Physically, the more relevant quantity 
is the ratio of the weight factors for any two different energy states as seen by the local observer.  Let us consider 
two energy states seen by the observer whose energies are removed from each other by an amount $\delta E$.  Then the 
ratio or the relative weight of the higher energy state would be  $\exp[-\beta_s\delta E]$. The smallest gap in energy 
that can be resolved by the observer would be of the order of $\frac{1}{\Lambda L}$. This gives the relative weight as 
$\exp[-\frac{\beta}{L}]$ or $\exp[-\frac{1}{aL}]$. While $aL\ll1$, it is not vanishingly small and it is independent of the factor 
$\frac{1}{\Lambda}$. 

\section{Discussion}
Jacobson's insight suggests that the Einstein equation could be a consequence of thermodynamics at a local 
level. This, however, works only in the local frame of those observers, who are accelerating at a
rate 
much higher than the scale set by the curvature of the spacetime at that point. Here we explore the possibility 
of extending this viewpoint to include observers, who may accelerate locally at a rate much less than the scale set by the scale of the 
curvature of the spacetime at that point. This necessitates a key change in the way the thermodynamic 
formulation is set up. In the case Jacobson considers, the Rindler horizon constitutes a part of the boundary 
of the local patch containing the observer. The energy passing through the horizon is then inaccessible to 
the accelerating observer and can be interpreted as the amount of energy that gets converted into heat in the frame of that 
observer. However, if the acceleration of the observer is not large compared to the scale set by the curvature 
of the spacetime, then the Rindler horizon does not constitute any part of the boundary of the  local patch 
containing the accelerating observer. In this case then, it is not possible to interpret the energy 
passing through the horizon as the heat energy, as seen by the observer. For a thermodynamic derivation of 
the Einstein equation to be possible in such cases, one has to assume that energy is being converted into 
heat even though no horizon exists. This forces 
one to adopt a more general proposal for determining the amount of energy that should be converted to heat 
in the local frame of the accelerating observer. By now, it has been quite convincingly demonstrated\cite{Paddy}
\cite{Paddy1}, \cite{Paddy2} that 
the null surfaces have some properties that are closely related to the thermodynamics of a system. 
This suggests the possibility, that for the locally accelerating observer, some fraction of the matter-energy
passing through the null surfaces is converted into heat. Apart from this, it is also 
assumed that the amount of heat generated in this way, depends on the 
non-inertial motion of the observer. This links the change in the timelike Killing vector 
across the null boundaries of the local patch to the amount of energy converted into heat. It allows 
one to view 
heat conversion in the case of matter falling through the horizon as a special case of a more general 
process of gravitational conversion of matter-energy into heat energy. This is the main new result of this 
paper.

It is interesting to make an estimate of the fraction of matter-energy that gets converted into heat 
according to this proposal. For observers locally accelerating at a rate much smaller than the scale set by the 
curvature of the spacetime, this factor is in fact quite small. A rough estimate of this conversion factor 
can be made by comparing the heat flux through the locally constructed two dimensional surface perpendicular 
to $n_\mu$ with the total energy flux through that same surface. The ratio of these two fluxes turns out to 
be of the order of $\sim an$. As has been argued earlier in this paper, $n\ll \frac{1}{\sqrt{Riemann}}$ 
and  $an\ll1$, for not so large values of the acceleration. For the local observer, the more relevant quantity 
is the fraction of the matter-energy that 
gets converted into heat per unit length of the local patch. This is proportional to the acceleration of 
that observer. In the rescaled coordinates of the localised observer, this acceleration is given by $a_s$. 
This is going to be very small due to the presence of the rescaling factor $\frac{1}{\Lambda}$. So the amount 
of heat generated would be rather tiny for the locally accelerating observe to measure. In the limit 
of $\Lambda$ going to infinity, the conversion factor goes to zero.

Another important feature of the heat generated in this way is that the amount of increase in heat as 
perceived by the local observer would depend 
on the curvature of the spacetime. This becomes clear if we look at the factor of conversion $a_s$ more 
closely. The factor that matters most here is $\frac{1}{\Lambda}$. Now, $\Lambda$ is of the order of the 
ratio of the size of the local patch to the smallest possible lengthscale of the localised wave modes that the locally 
accelerating observer could measure. If we denote the length of the size of the local patch by $L$, then 
$\Lambda \sim\frac{L}{u_{n_s}}$, where $u_{n_s}$ is the unit of length used by the local observer and it is of the 
same order as that of the lengthscale of the localised wave modes. Now if we assume that the magnitude of 
$u_{n_s}$ remains roughly of the same order, then the dependence of $\Lambda$ on the curvature can be expressed 
as $\Lambda \ll\frac{1}{(\sqrt{Riemann})u_{n_s}}$. This means that the conversion factor is of the order 
of $\frac{1}{\Lambda}\gg(\sqrt{Riemann})u_{n_s}$. So, the observer accelerating locally would find the amount of the energy that 
is being converted into heat to be greater, if the components of the Riemann curvature tensor at that point in 
spacetime is greater in magnitude. This suggests that the local observer perceives the amount of heat 
generated by gravity to be greater, when gravity is stronger.

It is also important to note that the way $ds$ was calculated for the causal diamond $\mathscr{D}$, 
tells us only about the total entropy increase. But it does not tell us how $ds$ is changing with 
time within the local patch. This is not surprising since we solved the Raychaudhury equation 
with a different boundary condition instead of the standard 
teleological condition used to prove the second law for the horizons. It is not clear at this moment,
how one can prove some version of the second law of thermodynamics in this case.

Finally, we also demonstrate here that a locally accelerating observer encounters a thermal environment with 
a temperature proportional to the value of her acceleration. This holds provided one is observing local 
physics only. This is reminiscent of the Unruh effect and also technically the calculation is very similar. 
However, there is a very important difference. In the case considered here, the locally accelerating 
observer does not observe thermal radiation at the temperature $a_s$ for the obvious reason
that the lengthscale corresponding to this temperature is much greater than the size of the local patch 
containing her. But any detector in her rest frame, interacting with much more localised modes 
of the radiation, would detect a thermal environment, The energy distribution in this case is found 
to be given by the Maxwell-Boltzmann distribution. This also means that the state seen by the observer need not be a thermal  
one. This is different from the Unruh effect, in which case the state seen by the Unruh observer is thermal.
But in the case dealt with here, it is not possible for local observers to stay within the local 
patch and  also determine whether the state seen by them is thermal or not. All they can do is determine 
the weights of the different energy-states( these are different from the quantum state of the field) and 
that comes out to be $\exp[-\beta_sE]$. 
This justifies the assignment of a temperature proportional to $a$ in the locally accelerating 
frame. A point can be made that 
how much significance should one attach to this result given that the state seen by the local observer may 
not be thermal. The point of view taken here is that the environment is indeed thermal as far as the local 
observers are concerned. Also the fact that it is possible to derive the Einstein equation from thermodynamic 
relations in this case lends some support to it. Taken together, they give credence to the view that for 
any locally accelerating observer,  the dynamics of gravity can be viewed as a thermodynamic process.

The temperature measured by the observer goes to zero as the acceleration $a$ vanishes. The observer 
for whom $a=0$, is falling freely in that local patch. So, it is possible for 
any observer who is not falling freely to view the Einstein equation locally as a manifestation of 
thermodynamic changes in a system. In this sense, the result presented here is a generalisation of 
Jacobson's derivation of Einstein equation from thermodynamics to the case of any non-freely falling 
observer. A freely falling observer in that same local patch does not detect any temperature or the 
heating up of null surfaces 
by gravity. This is similar to the difference that exists between a Rindler-like observer and an inertial 
observer or between an observer freely falling through the black hole horizon vis-a-vis one who remains 
outside. A final remark is in order here. Due to the equivalence principle, the above considerations 
would apply to any observer in a locally constant gravitational field as well.

\section*{Acknowledgements}
I am greatly indebted to Sudipta Sarkar for helping me understand many issues in horizon thermodynamics and 
also in Jacobson's paper. I am also grateful to Ghanashyam Date for the long discussions and the detailed comments,
helpful suggestions and clarifications he made on the third section of the paper. Alok Laddha has 
also been very helpful with several comments and important suggestions. I would also like to thank 
Kinjalk Lochan and 
Satyabrata Sahu for discussions, which encouraged me to look at this problem. A P Balachandran had made some helpful 
comments at an earlier stage. Thanks are also due to Gaurav Narain and Saurav Gupta for helpful discussions and 
suggestions. Last, but not the least, I am grateful to Sudipto PalChowdhury for keeping his patience throughout our 
discussions related to this work.

.

\end{document}